\documentclass[twocolumn,amssymb,showpacs,floatfix]{revtex4}
\usepackage{graphicx}
\usepackage{amssymb}
\begin{document}
\title{
Extended Standard Map with Spatio-Temporal Asymmetry
}

\author
{
Taksu Cheon${ }^{1}$, 
Pavel Exner${ }^{2,3}$, and  
Petr {\v S}eba${ }^{3,4}$
\email
{emails: taksu.cheon@kochi-tech.ac.jp, exner@ujf.cz, petr.seba@uhk.cz}
}
%
\affiliation{
 ${ }^1$
 Laboratory of Physics, Kochi University of Technology,
Tosa Yamada, Kochi~782-8502, Japan\\
 ${ }^2$ 
Nuclear Physics Institute, Czech Academy of Sciences,
25068~\v{R}e\v{z} u Prahy, Czech Republic\\ 
 ${ }^3$ 
Doppler Institute, Czech Technical University,
B\v{r}ehov\'{a} 7, 11519~Praha, Czech Republic\\ 
 ${ }^4$ 
Department of Physics, University of Hradec Kralove,
50003~Hradec Kralove, Czech Republic
}
%
%
\date{March 3, 2002}
%

%
%
\begin{abstract}
We analyze the transport properties of a set of symmetry-breaking 
extensions
of the  Chirikov--Taylor Map.
The spatial and temporal asymmetries result in the loss of 
periodicity in momentum direction in the phase space dynamics, 
enabling the asymmetric diffusion which is the origin of the 
unidirectional  motion.
The simplicity of the model makes the calculation of the global dynamical 
properties of the system feasible both in phase space and in
controlling-parameter space.
We present the results of numerical experiments which 
show the intricate dependence of the asymmetric diffusion 
to the controlling parameters. 

\vspace*{0.5mm}
\noindent KEYWORDS:
  deterministic diffusion,
  chaotic transport,
  molecular motor
\end{abstract}
\pacs{ 5.40.Fb, 5.45.-Gg, 5.60.Cd, 87.15.Aa}
\maketitle
%
%
\section{Introduction}
The motion of a particle in temporally and spatially periodic 
potential has attracted renewed attention as a model of 
molecular motors \cite{JA97}.  In order to make it a motor, 
that generates unidirectional motion out of 
periodic motion, the potential has to be a ``ratchet'',
{\it i.e.} its reflection symmetry in both spatial and temporal 
direction has to be broken, a fact generally known as a 
Curie's theorem.
Traditionally, the friction has supplied the source of 
temporal symmetry breaking \cite{BH94,JK96,PU00,PA00}.
There are several recent works \cite{FY00, SO01} in which 
the frictionless ``inertial'' regime is studied
where the time-asymmetric perturbation is the
driving force of the unidirectional motion.
Through these studies, it has become evident that the characeristics
of the phase space plays a dicisive role. The exploration of the 
full phase space of a dynamical system, however, 
requires considerable computational resources.
This naturally leads to models of ratchet dynamics in 
which the differential equation governing the time evolution is 
reducible to the discrete difference equations, 
namely the {\it map} \cite{SO01}. 
One would then like to have a model of a system with ratchet dynamics,
which is free from {\it ad hoc} assumptions and parameters that tend
to obscure the underlying simple physics.

In this article, we propose a simple model of a particle motion
under temporally and spatially periodic potential whose 
key feature is in that the obtained descrete map is
a natural extention of the Chirikov-Taylor map, also known as 
the standard map \cite{CH79}  whose dynamical properties have been
thouroughly analyzed.
The unidirectional motion of the system is understood
in the context of the deterministic diffusion found in the standard map. 
Also, as a practical benefit of the discrete evolution of the map dynamics,
the computational burden is substantially reduced, 
and it becomes possible to study, for the first time,
the transportation properties of a system 
in a broad region in the controlling parameter space.
A very intricate dependence of the unidirectional motion on the
controlling parameter is revealed though a numerical investigation.

%
\section{Extended Standard Map: Frictionless case}
We consider the motion of a classical particle described 
by a time-periodic Hamiltonian with instantaneous potential action, 
or $\delta$-function kick, which has 
space periodic part and alternating-current type 
driving motion;
\begin{eqnarray}
\label{eq1}
H={{p^2} \over 2}+\sum\limits_{n=-\infty }^\infty  
{\left[ {u(x)+xAs_n} \right]
\delta (t-nT\{ 1+\varepsilon s_n\} )} ,
\end{eqnarray}
where, for now, the trapping potential $u(x)$ is set to be symmetric 
and given by
\begin{eqnarray}
\label{eq2}
u(x)=-K \cos x ,
\end{eqnarray}
and $s_n\equiv (-1)^n$ is the parity of the iteration number $n$.
The evolution of the system is described by
\begin{eqnarray}
\label{eq3}
\dot x &=& p
\\
\label{eq4}
\dot p &=& -\sum\limits_{n=-\infty }^\infty  
{\left[ {u'(x)+As_n} \right]
\delta (t-nT\{ 1+\varepsilon s_n\} )} ,
\end{eqnarray}
One can 
integrate the motion in between the $\delta$-function kicks and 
obtain the map which describe the evolution of 
position $x_n$ and momentum $p_n$ at discrete time 
$t_n=nT(1+\varepsilon s_n)-0$;
\begin{eqnarray}
\label{eq5} 
x_{n+1} &=& x_n+p_{n+1}(1+\varepsilon s_n)
\\
\label{eq6}
p_{n+1} &=& p_n-K\sin x_n-As_n .
\end{eqnarray}
This set of equations can be regarded as a three-parameter
extension of the standard map \cite{CH01}. 
The parameter $K$, which is present in original standard map
is the strength of the 
periodic potential that traps the particles 
in each periodic location.  The parameter $A$ is the 
amplitude of alternating swing motion, and $\varepsilon$ is 
the measure of its time asymmetry. 
As with the case of the standard map,
the dynamical properties of the system is studied through
the analysis of phase space \cite{RE91}.  
The transport property of standard map, 
$A$ = $\varepsilon$ = 0, is characterized in terms of phase space average 
of functions of position and momentum $x$ and $p$.  It is well known 
that the direct averages $\left\langle x\right\rangle$
and $\left\langle p\right\rangle$ are zeroas a direct
result of the fact that the standard map is both time-reversal 
and space-reflectional symmetric. 
However the mean square $\left\langle p^2\right\rangle$  
can be non zero when one has the chaotic motion unbounded 
by KAM tori in $p$-direction. 
This happens with sufficiently large value of $K$.
One then has deterministic diffusion through chaotic motion which is 
characterized by the random-walk behavior, 
$\left\langle x^2\right\rangle$ $\propto$ $t$.
In the extended standard map (\ref{eq5})-(\ref{eq6}),
a unidirectional  motion can be realized if one has 
by the non-zero value for the $\left\langle p\right\rangle$ 
itself in the diffusive motion. 

The reflection symmetries play an essential role in the existence 
(or non-existence) of the unidirectional motion \cite{FY00}.
One can rewrite  (\ref{eq5})-(\ref{eq6}) as
\begin{eqnarray}
\label{eq7} 
-x_{n+1} &=& -x_n-p_{n+1}(1+\varepsilon s_n)
\\
\label{eq8}
-p_{n+1} &=& -p_n-K\sin{(-x_n)}+As_n.
\end{eqnarray}
These become identical in form to  (\ref{eq5})-(\ref{eq6}) themselves
in case of $A$ = 0. 
Therefore, in this case, one has
\begin{eqnarray}
\label{eq9} 
\left\langle {x_n} \right\rangle 
&=& \left\langle {-x_n} \right\rangle = 0 ,
\\ \nonumber
\left\langle {p_n} \right\rangle 
&=& \left\langle {-p_n} \right\rangle = 0
\ \ \ \ \ \ 
({\rm for\ } A = 0 ) ,
\end{eqnarray}
which means that there cannot be  any directed transport.

We now define $\tilde x_n \equiv x_n$ 
and $\tilde p_n \equiv p_{n+1}$, obtain a reverse map 
\begin{eqnarray}
\label{eq10} 
\tilde x_{n-1} &=& \tilde x_n-\tilde p_{n-1}(1-\varepsilon s_n)
\\
\label{eq11}
-\tilde p_{n-1} &=& -\tilde p_n-u(\tilde x_n)-As_n .
\end{eqnarray}
Therefore, if one sets $\varepsilon$ = 0, the map 
for $\{x_n, p_n\}$ and $\{ \tilde x_{-n}, -\tilde p_{-n}\}$ are 
identical, and one has
\begin{eqnarray}
\label{eq12} 
\left\langle {-p} \right\rangle 
=\left\langle {-\tilde p} \right\rangle 
=\left\langle p \right\rangle =0
\ \ \ \ \ 
({\rm for\ }\varepsilon = 0)
\end{eqnarray}
Therefore, one needs both $A \ne 0$  
and $\varepsilon \ne 0$  to have 
directed motion.  
It is important to note that the  map (\ref{eq5})-(\ref{eq6}) in fact atcs
differently for odd and ven $n$ because of the existence of $s_n$ term.
It is advantageous to consider the phase spaces for odd $n$ and even $n$
separately.  The reflection symmetry between  (\ref{eq5})-(\ref{eq6}) and
 (\ref{eq7})-(\ref{eq8}) should then be interpreted as one between two 
phase space profiles each representing odd $n$ and even $n$. 
%
\begin{figure}
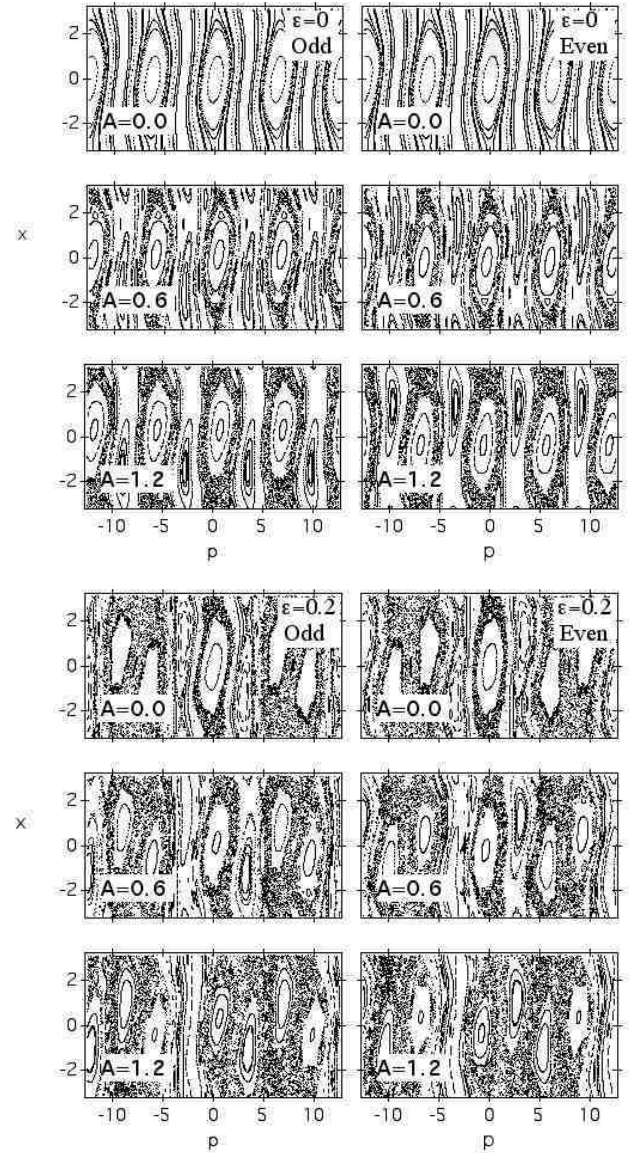

\includegraphics[width=90mm]{ff1a.epsf}
\includegraphics[width=90mm]{ff1c.epsf}
\caption{
Phase space profiles of the map Eqs. (5)-(6) for even numbers for $n$.
The strength of periodic potential is set to be $K = 0.6$.
The temporal asymmetry parameter $\varepsilon$ and amplitude of 
alternating swing $A$ are varied.  Left (right) column is for odd (even) $n$.
 }
\end{figure}
%

For the standard map, $A =$ $\varepsilon = 0$,
as is well known, the map is $2\pi$-periodic in both
$x$ and $p$ directions.
The periodicity for $x$ direction is kept intact 
for non-zero $A$ and $\varepsilon$:  when one has a map
$(x_n, p_n)$ $\rightarrow$ 
$(x_{n+1}, p_{n+1})$, one also has 
\begin{eqnarray}
\label{eq13} 
(x_n+2\pi , p_n)
\rightarrow
(x_{n+1}+2\pi , p_{n+1}) .
\end{eqnarray}
On the other hand, original $2\pi$ periodicty for 
$p$ - direction is lost when non-zero $\varepsilon$ is present.
If, however, $\varepsilon$ is a rational number, $M/N$ 
where $M$ and $N$ are mutually incomensurate positive integers, 
a new $2N\pi$-periodicty for $p$-direction emerges, because 
in this case, one has
\begin{eqnarray}
\label{eq14} 
(x_n , p_n+2 N \pi)
&\rightarrow&
\\ \nonumber
(x_{n+1}&+2&(N \pm M) \pi , p_{n+1}+2N\pi)
\\ \nonumber
& &
\ \ \ \ \ \
({\rm for\ } \varepsilon = M/N; N > M > 0) .
\end{eqnarray}
%
\begin{figure}
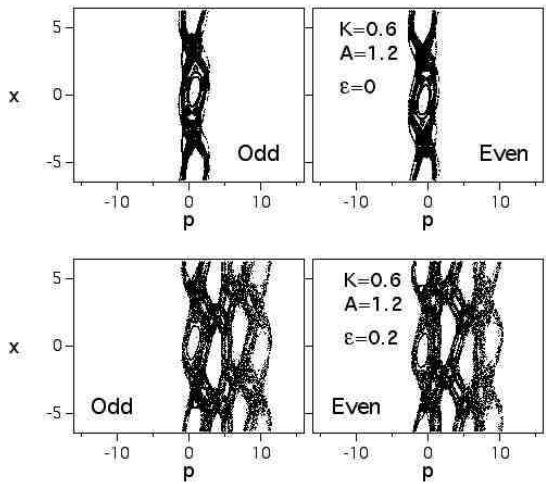

\includegraphics[width=80mm]{ff2a.epsf}
\includegraphics[width=80mm]{ff2b.epsf}
\caption{
Phase space orbits of the map (5)-(6) 
for $\varepsilon = 0$ (top) and $\varepsilon = 0.2$ (bottom)
started from a set of random initial
conditions confined in the region $|p|$ $<$ $\pi/10$ and
$|x|$ $<$ $\pi/2$. 
The strength parameters are set to be $A = 1.2$ and $K = 0.6$.
For the case of $\varepsilon = 0$, 
one has
$\left\langle p \right\rangle_{odd} +
  \left\langle p \right\rangle_{even} $ $= 0$.
}
\end{figure}

The change of the $p$-periodicity has a direct impact on 
the transportation properties of the extended standard map.
The $2N\pi$-periodicity in $p$-direction 
for  a rational number $\varepsilon = M/N$ (where $M$ and $N$
are two mutually incommensurate natural numbers)
means that the KAM-tori
(when they exit) appear at least once in $2N \pi$ period.
Thus the chaotic diffusion is bounded within the region 
in between two adjacent tori.  
This fact gives an estimate for the average 
momentum $\left\langle p \right\rangle$ within a
region that include $p = 0$ in the form
\begin{eqnarray}
\label{eq15}
\left\langle p \right\rangle  \le N\pi 
\ \ \ \ \
({\rm for\ } \varepsilon = M/N; N > M >0).
\end{eqnarray}
Note that only the upperbound is obtained since the 
exact location of the two adjacent KAM tori near $p=0$ 
cannot be determined {\it a priori}. 
This is evident when one considers $\varepsilon = 0$ case,
in which the map is $2\pi$-periodic in $p$-direction,
and a similar estimate to above gives 
$\left\langle p \right\rangle$  $\le$ $\pi$, while,
as we have lernt in (\ref{eq12}), it is zero because KAM tori is placed
symmetrically around $p=0$. 
%
\begin{figure}
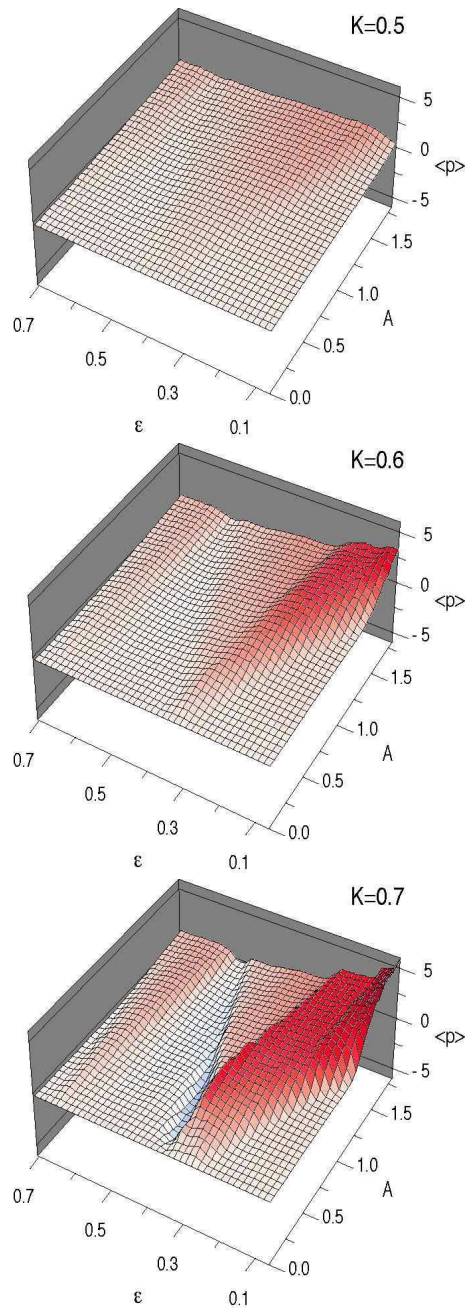

\includegraphics[width=60mm]{ff3a.epsf}
\includegraphics[width=60mm]{ff3b.epsf}
\includegraphics[width=60mm]{ff3c.epsf}
\caption{
A measure of directed motion $\left\langle p \right\rangle$
as the function of asymmetry parameter $\varepsilon$ and 
the strength of swing motion $A$ 
for the extended standard map with asymmetric kick (5)-(6) 
}
\end{figure}
%

The situation is illustrated in Fig.1 where we draw phase 
space profiles for the frictionless extende standard map with 
various $\varepsilon$ and $A$. The strength parameter is chosen 
to be $K$ $=0.6$ in these examples.
The effect of the breaking of phase space symmetries on 
the transportation properties of the system becomes evident
by the inspection of Fig. 2, where we plot the portion of 
the phase space profile that is reached from a set of initial
conditions clustered around in the vicinity of $p = 0$.
With the onset of non-zero $A$ and non-zero $\varepsilon$ terms,
the average momenta in the chaotic region around $p = 0$ 
is shifted from zero, and the directed tranport is realized.
This is essentially in accordance with 
the conclusion of Flach et al. \cite{FY00}, 
in which they identify this phenomenon
asymmetrical Levy flight.  
We are now ready to ready to look at the directed transport
not just in one parameter set, but in a global region in 
the parameter space.  In Fig. 3, we plot the average momentum
$\left\langle p \right\rangle$ of the chaotic band that include
$p = 0$ as the function of swing parameter $A$ and time asymmetry
$\varepsilon$. 
The coupling strength $K$
is taken to be $K = 0.5$ (top), $K = 0.6$ (middle) and $K = 0.7$ (bottom).
We start from a ensemble of 800 random initial configuration
satisfying  $|p|$ $<$ $\pi/10$ and $|x|$ $<$ $\pi/2$.  
We symmetrize the ensemble in terms of $p-$ and $x-$ to guarantee 
$\left\langle p \right\rangle = 0$ exactly at the start.
Each configuration is
evolved with (5)-(5) for 12,000 times, and the momentum $p$ is averaged
over that period and over the ensemble to obtain  $\left\langle p \right\rangle$
for a given $A$ and $\varepsilon$.  
Calculation on the grid points of $48 \times 42$ in the space 
of $(A, \varepsilon)$ is used for each of the three figures.

From these figures, it appears that
the direction of the asymmetric transport 
(positivity or negativity) is controllable, having
step-like structure of increasing $\left\langle p \right\rangle$ 
with decreasing $\varepsilon$, which is in broad accordance 
to our arguments leading to (\ref{eq15}). 
An intriguing feature is the direction-reversal of
$\left\langle p \right\rangle$ with parameter variation, whose 
occurence is not at all obvious in the outset.
Note also that 
the speed of unidirectional transport can be rather large: 
$\left\langle p \right\rangle$
 is almost $2\pi$ in the peak region, which means 
that the particle moves one spatial period in one 
temporal period.    

%
\section{Ratchet map with friction}

Our model can be extended to incorporate the ratchet dynamics
with damping through the addition of friction.
In place of eqs. (\ref{eq5})-(\ref{eq6}), we set
\begin{eqnarray}
\label{eq16}
\dot x &=& p
\\
\label{eq17}
\dot p &=& -\sum\limits_{n=-\infty }^\infty  
{\left[ {u'_{AS}(x)+As_n} \right]}
\\ \nonumber
& &
\times
{\delta (t-nT\{ 1+\varepsilon s_n\} )}
-\gamma p
\end{eqnarray}
where we now add spatially asymmetric term to 
the trapping potential as
\begin{eqnarray}
\label{eq18}
u'_{AS}(x)=K\{ \sin x_n+2\mu \sin (2x_n+\delta )\}.
\end{eqnarray}
The map obtained from these is written as
\begin{eqnarray}
\label{eq19}
x_{n+1} &=& x_n+p_{n+1}
(e^{\gamma (1+\varepsilon s_n)}-1)/\gamma 
\\
\label{eq20}
p_{n+1} &=& 
\left[ {p_n - u'_{AS}(x_n)-As_n} \right]
e^{-\gamma (1+\varepsilon s_n)}
\end{eqnarray}
\begin{figure}
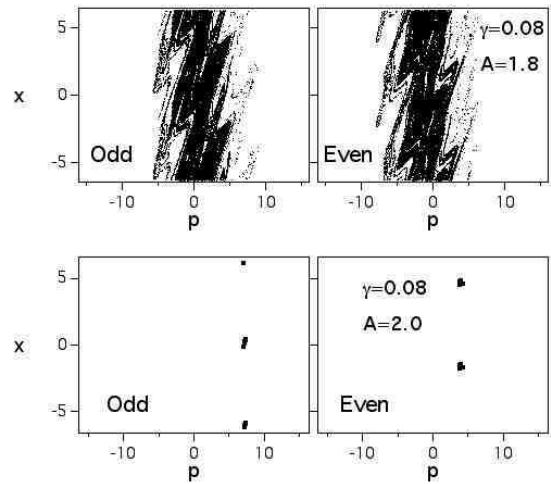

\includegraphics[width=80mm]{ff4a.epsf}
\includegraphics[width=80mm]{ff4b.epsf}
\caption{
Phase space orbit of map (21)-(22) after 10000 iteration
showing the starnge attractor for $A=1.8$ (top figures), 
and the periodic attractor $A=2.0$ (bottom figures).
Other parameters are set to be $\mu = 0.2$ and $\delta=0.3$. 
}
\end{figure}
%

There are now six parameters in the model:
$K$, $A$, $\mu$, $\delta$, $\gamma$, and $\varepsilon$.  
We reduce this number by setting $\varepsilon$ $=$ $0$.
Since the temporal asymmetry is already broken by
$\gamma$ alone, we do not expect to loose out much. 
We now obtain
\begin{eqnarray}
\label{eq21}
x_{n+1} &=& x_n+p_{n+1}
(e^{\gamma}-1)/\gamma 
\\
\label{eq22}
p_{n+1} &=& 
\left[ {p_n - u'_{AS}(x_n)-As_n} \right]
e^{-\gamma}
\end{eqnarray}
%
\begin{figure}
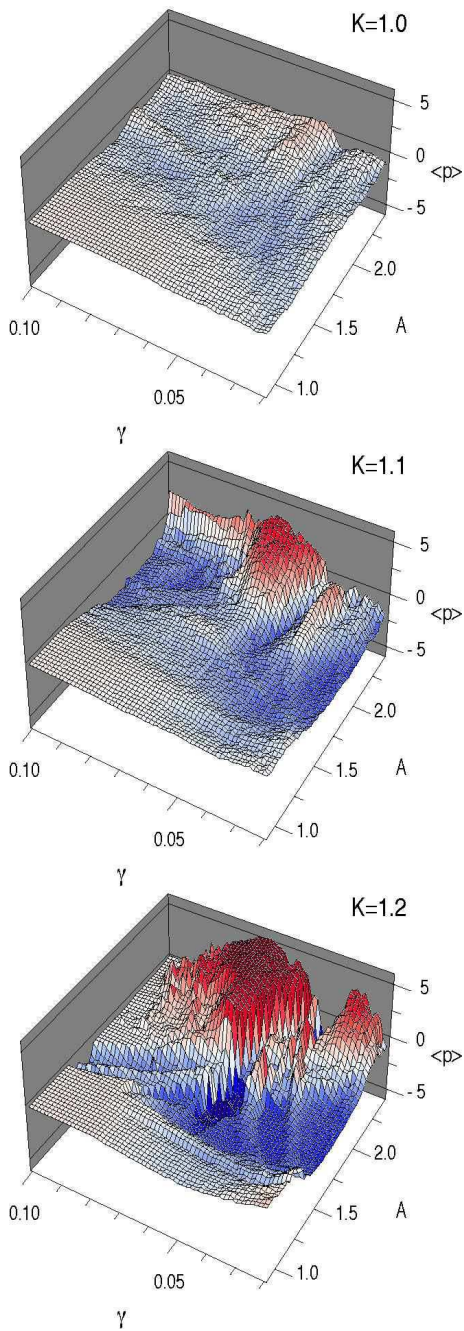

\includegraphics[width=60mm]{ff5a.epsf}
\includegraphics[width=60mm]{ff5b.epsf}
\includegraphics[width=60mm]{ff5c.epsf}
\caption{
The average $\left\langle p \right\rangle$
as the function of $\gamma$ and $A$ for the extended standard map 
with damping (21)-(22) 
The parameters controling the spatial asymmetry 
are set to be $\mu = 0.2$ and $\delta=0.3$. 
}
\end{figure}
%
%
As in the frictionless case, we consider the reflection symmetry of the
map (\ref{eq21})-(\ref{eq22}).
With with the diffinition $\tilde x_n \equiv x_n$ 
and $\tilde p_n \equiv p_{n+1}$, one obtains a reverse map 
\begin{eqnarray}
\label{eq23} 
\tilde x_{n-1} &=& \tilde x_n-\tilde p_{n-1}(\gamma-1)/\gamma
\\
\label{eq24}
-\tilde p_{n-1} &=& -\tilde p_n e^{\gamma}
                   +[-u'_{AS}(\tilde x_n)-As_n]e^{-\gamma}
\end{eqnarray}
In the elastic limit $\gamma$ $\to$ $0$, the map 
for $\{x_n, p_n\}$ and $\{ \tilde x_{-n}, - \tilde p_{-n}\}$ are 
identical, and one has
\begin{eqnarray}
\label{eq25} 
\left\langle {-p} \right\rangle 
=\left\langle {-\tilde p} \right\rangle 
=\left\langle p \right\rangle =0
\ \ \ \ \ 
({\rm for\ }\gamma \to 0)
\end{eqnarray}
irrespective to the symmetry property of $u_{AS}$.
This result, puzzling in first inspection, is of course
in accordance with Curie's theorem.
If, on the other hand, we have $A = 0$, the motion
is overdamped, and in sufficient steps, it will come
to stand still, namely,
\begin{eqnarray}
\label{eq26}
x=x_0
\ \ \ 
( u'_{AS}(x_0) = 0 ),
\ \ \
p=0
\ \ \ \ \ 
({\rm for\ }A = 0)
\end{eqnarray}
%
With friction, after sufficiently large steps $n$, all trajectories 
fall into the atractors. They can be either stable attractors with finite
period, or the strange attractors with fractal dimension depending on the
parameter value \cite{OT93}.  
Two examples from each case are shown as the top and bottom 
figures in Figs. 4.
As is well known, these two cases alternate with minute variation of 
parameters.  
If one has $\mu =$ $\delta = 0$,  the shape of attractor respects 
the spatial reflection symmetry, when both even and odd $n$
are considered. 
Therefore, one can only have $\left\langle p \right\rangle$ in this case.
On the other hand,
as long as $\mu \ne 0$ and $\delta \ne 0$, one has no spatial 
mirror symmetry in the phase space, and one should have
 $\left\langle p \right\rangle \ne 0$ in general.  
It is expected 
that the value of $\left\langle p \right\rangle$ 
can vary wildly as fuctions of parameter because
of the afore-mntioned alternation of periodic and strange attractors
with minute change of he parameter.

The transport property of the extended standard map with
friction can be overviewd in Fig. 5, in which we show the three-dimensional
plot of $\left\langle p \right\rangle$ as functions of two controlling 
parameters $A$ and $\gamma$.  The spatial asymmetry parameters 
are fixed to be $\mu = 0.2$ and $\delta = 0.3$.  The coupling strength $K$
is varied from $K = 1.0$ to $K = 1.2$ from the top to bottom.  As before, 
we start from a $p-$ and $x-$symmetrized ensemble of 1600 initial configuration
satisfying  $|p|$ $<$ $\pi/10$ and $|x|$ $<$ $\pi/2$.  Each configuration is
evolved with (\ref{eq21})-(\ref{eq22}) for 40,000 times, and the 
momentum $p$ is averaged
over that period and over the ensemble to obtain  $\left\langle p \right\rangle$
for a given $A$ and $\varepsilon$.  
The grid points of $64 \times 64$ in the parameter 
space of $(A, \gamma)$ in the range $A \in [0.8, 2.8]$ 
and $\gamma \in [0.02, 0.10]$
are used for each of the three figures.

One can see from these figures, that the 
overdamped motion in ratchet potential shows directed transport, but
is less controllable than the inertial counterpart, as expected from our 
preceeding arguments.  At the moment, we do not have any clue as to the
physical origin of the intriguing structure found in Fig. 5.  However, we feel that
the simplicity of our model  (\ref{eq21})-(\ref{eq22}) should yield to a 
mathematical analysis  eventually to give us such clue.

%
\section{Summary and Prospects}
In this work,  an extended standard map
which has spatial and temporal asymmetry is formulated.
We have outlined its phase space properties, and have
studied the unidirectional diffusion, 
which is thought to be relevant for molecular and biological motors.
There exist several  studies of great mathematical rigor on 
the standard map \cite{GR79,MK87}. 
Our extended standard map may hopefully become a subject of similar
research.
One immediate extension of the current model is to introduce the
randomness.  Namely, one may consider $\varepsilon$ not as a fixed
quantity, but a fluctuating number for each $n$.  This is a simple way to 
include the thermal fluctuation into the model.
Lastly, the study of quantum version of the current model may be of 
great interest.  This is particularly so, because the quantum version of 
standard map  \cite{CC79,FG82} is again a standard tool in the study of
quantum chaotic diffusion.

%
\subsection*{Acknowledments}
TC expresses
his thanks to Prof. Izumi Tsutsui 
for helpful dicsussions. He also thanks members of the Theory
Group of High Energy Accelerator Research Organization (KEK) 
at Tsukuba for the Hospitality extended to him during his stay.
This work has been
supported in part by the Grant-in-Aid (C) (No. 10640413) of the
Japanese Ministry of Science and Education, and
also by GAAS and the Czech Ministry of Education 
within the projects A1048101 and ME170.

%

%
%
\end{document}